# Spatiotemporal mode-locking in lasers with large modal dispersion

Yihang Ding,[1]† Xiaosheng Xiao,[2]*† Kewei Liu,[1] Shuzheng Fan,[2] Xiaoguang Zhang,[2] Changxi Yang[1]*

[1]State Key Laboratory of Precision Measurement Technology and Instruments, Department of Precision Instruments, Tsinghua University, Beijing 100084, China.

[2]State Key Laboratory of Information Photonics and Optical Communications, School of Electronic Engineering, Beijing University of Posts and Telecommunications, Beijing 100876, China.

***ABSTRACT*** Dissipative nonlinear wave dynamics have been investigated extensively in mode-locked lasers with single transverse-mode, whereas there are few studies related to three-dimensional nonlinear dynamics within lasers. Recently, spatiotemporal mode-locking (STML) was proposed in lasers with small modal (i.e., transverse-mode) dispersion, which has been considered to be critical for achieving STML in those cavities because the small dispersion can be easily balanced. Here, we demonstrate that STML can also be achieved in multimode lasers with much larger modal dispersion, where we find that the intracavity saturable absorber plays an important role for counteracting the large modal dispersion. Furthermore, we observe a new STML phenomenon of passive nonlinear auto-selection of single-mode mode-locking, resulting from the interaction between spatiotemporal saturable absorption and spatial gain competition. Our work significantly broadens the design possibilities for useful STML lasers thus making them much more accessible for applications, and extends the explorable parameter space of the novel dissipative spatiotemporal nonlinear dynamics that can be achieved in these lasers.

Mode-locked lasers based on single-mode fibers are popular platforms for studying the behavior of dissipative nonlinear wave dynamics [1], because they provide a simple way to perform nonlinear science research. Additionally, mode-locked fiber lasers are promising for many applications. However, single-mode fibers are challenged by the ever-increasing demand for higher-energy lasers, as well as for larger data capacities in the field of optical fiber communications [2,3]. The use of multimode fibers (MMFs) is the most obvious way to overcome these limitations due to the larger mode area and additional spatial degree of freedom of MMFs.

In the field of nonlinear science, a multimode optical system based on MMFs, combining both temporal and spatial characteristics, can provide a better test bed than a conventional system composed of single-mode fibers. There is great significance in studying nonlinear dynamics in 3-dimensional optical systems composed of MMFs, as these dynamics are related to important issues in physics in term of spatiotemporal dynamics. The various nonlinear spatiotemporal dynamics in passive graded-index MMFs have been investigated extensively recently, including studies of spatiotemporal instability [4-6], spatial beam cleaning [7-12], supercontinuum generation [12-17], and control of nonlinear multimode interactions [18,19]. In addition, nonlinear pulse propagation in amplifiers composed of active MMFs was studied [20-23]. However, very little work on MMF cavities with multimode nonlinear dynamics has been performed.

Recently, L. Wright et al. proposed spatiotemporal mode-locking (STML), i.e., both the longitudinal- and transverse-modes were locked simultaneously, in a cavity composed of graded-index MMFs [24], and the key factor that makes the STML possible in those MMF cavities was considered as the small modal dispersion of the graded-index MMFs. The small modal (i.e., spatial or transverse-mode) dispersion of graded-index MMFs is comparable to chromatic (i.e., longitudinal-mode) dispersion and these dispersions can be compensated for by strong intracavity spatial and spectral filtering in those cavities [24], allowing transverse-modes to travel together. According to this, large modal dispersion will pose a challenge for achieving STML in MMF cavities. It is interesting to know whether and how STML can be supported in MMF cavities with large modal dispersion.

The realization of STML inside an MMF cavity relies on a delicate balance among the intracavity linear and nonlinear effects, including modal and chromatic dispersions, inter- and intra-modal nonlinearities, filtering, gain and loss. The significant influence of cavity dispersion on the dissipative nonlinear dynamics has been widely observed and investigated in single-mode fiber lasers [1]. Similarly, in MMF cavities, different dispersion parameter will cause different dynamical process of self-organization and different nonlinear spatiotemporal phenomena. Therefore, if STML can be achieved in MMF cavities with large modal dispersion, a further important question is what the dynamical process of the spatiotemporal self-organization is.

As the counterpart of the graded-index fibers, step-index fibers usually exhibit much larger (ten times) modal dispersion (i.e., walk-off among different modes) ([25,26] and sections S3.1 of Supplemental Material [27]). Thus step-index MMFs can serve as an appropriate medium for our curious investigations. From an application perspective, the development of useful STML lasers are greatly limited by the rarity of graded-index multimode gain fibers, which are not commercially available, likely due to the complex distribution of the refractive index compared to that of step-index fibers. Quasi-single-mode gain fibers have been used in some cavities ([34-36] and the first cavity of [24]). Although these cavities allow for observing a variety of spatiotemporal effects (e.g., soliton molecule reported in our previous work [34]), they are severely restricted compared to full-multimode systems in terms of both the range of physics that can be observed and the potential performance of the lasers. Contrary to the rarity of active graded-index MMFs, the active step-index ones are easy to commercially obtain, despite the much larger modal dispersion.

In this Letter, we show how to realize STML by using active step-index MMFs with large modal dispersion, and report the observation of novel nonlinear spatiotemporal dynamics within the cavity. In these MMF cavities with large modal dispersion, the key to achieve STML is the effects of intracavity saturable absorber (SA). By optimizing the lengths of MMFs, the SA plays two critical roles. One is the basic function of SA, passively modulating the cavity loss, which is usually expected, and the other is balancing the large walk-off among transverse-modes, which is an additional effect achieved in this cavity. It is found that the spatiotemporal evolution of the multimode pulses is quite different from that inside the graded-index MMF cavity. Moreover, we observed an interesting nonlinear phenomenon of passive beam auto-selection, i.e., the transition from multimode to single-mode mode-locking state. We found evidence for this important regime in which the fast nonlinear effects of SA interact with the spatial gain competition to lead to the passive nonlinear auto-selection of single-mode mode-locking.

Numerical simulations are first conducted to analyze the spatiotemporal evolution of pulses in an MMF cavity with step-index gain fibers (detailed in sections S1-3 of [27]). The cavity used in the simulations is shown in Fig. 1(a). The light is amplified through propagation in an active step-index MMF and then coupled with a segment of passive graded-index MMF and transmitted through a lumped SA. Then a spectral filter is applied, and the light is injected back into the gain fiber. The spectral filter is usually used in mode-locking fiber lasers with normal dispersion to meet the requirement of pulse self-consistency [37]. The step-index MMF is the same as that in Fig. S1, and the parameters of both MMFs are in accordance with those of the fibers used in the following experiments [27].

With appropriate cavity parameters, we found that stable STML could be achieved in the MMF laser cavity with the active step-index MMF. Typical numerical result for the STML state is presented in Figs. 1(b-e) (see more in sections S2, S3 of [27] and Movie S1 [38]). The intercavity evolution of the output of two typical modes, modes 1 and 6, over 80 roundtrips in the cavity is shown in Fig. 1(b), starting from small pulses. The pulses reach a steady operation state quickly. The intracavity pulse evolution in the final roundtrip is plotted in Fig. 1(c), where the walk-off between two typical modes, modes 1 and 6, is presented. The walk-off between the two modes constantly increases in the MMFs. Notably, the dramatic decrease in walk-off by SA indicates that SA plays an important role in pulling the modes together to achieve self-consistency and realizing STML in the cavity. The mode-resolved temporal output is plotted in Fig. 1(d), and Fig. 1(e) shows the corresponding spatiotemporal intensity of the output pulse. Overall, the combination effects of intracavity spatial filtering, spectral filtering and SA are strong enough to counteract the large modal dispersion. Among them, the SA takes a great proportion in balancing the dispersion, which is the key to let STML work in the MMF cavities with large modal dispersion.

Note that in the graded-index fiber, the movement speed of mode 6 is much larger than the linear velocity calculated from modal dispersion, which is indicated by the black dashed line in Fig. 1(c). This larger walk-off is induced by nonlinear multimode interactions, which is greatly affected by the large modal dispersion in step-index fiber (see details in section S3.3 of [27]). Comparing Fig. 1(c) to the evolution in a similar cavity but with small modal dispersion shown in Fig. S3 [27], it shows that the large modal dispersion results in quite different evolutions.

Based on the simulations, we established an MMF laser cavity with a step-index gain fiber accordingly (see Fig. S5 in [27]). Nonlinear polarization rotation was used to establish an artificial and ultrafast SA [39]. Figure 2 records the gradual transition to STML in the MMF cavity with increasing pump power (recorded every 0.1 W). Once the pump power reaches 2.5 W, STML can be easily self-started by adjusting the pump power only. The spectra and corresponding beam profiles at different pump powers are depicted in Fig. 2(d-k). As the pump power increases through the thresholds (around 2.5/3.7 W), sudden changes take place in the beam profile, as well as the temporal and spectral outputs, as the operation state transits between multimode continuous-wave lasing and mode-locking, indicating that STML occurs. More proofs of STML are given in section S5 of the supplementary material [27].

To determine which cavity can support STML, we experimentally tested many cavities with various lengths of the MMFs (see section S5.3 of [27]). In general, a long step-index fiber induces large walk-off among the transverse-modes, which is difficult to compensate for. Additionally, a long passive graded-index fiber induces a strong SA effect through the accumulation of nonlinear polarization rotation, which aids in the formation of STML. Therefore, a cavity configuration with a short step-index gain fiber plus a relatively long graded-index fiber is preferred for achieving STML.

Furthermore, we experimentally observed a nonlinear phenomenon of beam auto-selection mode-locking, i.e., a multimode STML state transiting to a mode-locking state with a single transverse-mode by adjusting the orientation of the intracavity waveplate only. A typical transition is given in Fig. 3. With appropriate optical coupling state, waveplate sets, pump power, etc., in the cavity, stable multimode STML is achieved. The beam profile of the output shown in Fig. 3(a) indicates that it is multimode. Then, by rotating the waveplate carefully, the multimode STML state transitions to another mode-locking state, as shown in Fig. 3(d-g). Compared to the old state, the beam profile and spectrum of the new state change, while the repetition rate remains the same.

The output beam profile of the new state, measured in the near field [Fig. 3(d)], contains two symmetric lobes. By the assumption of the $LP_{11}$ profile, one can conduct that the field can be described by a Laguerre Gauss distribution [11]. According to the mathematical properties of the Fourier transform of the Laguerre Gauss function, the far field of the $LP_{11}$ mode is a Laguerre Gauss distribution as well. The far-field beam profile is recorded in Fig. 3(e), with a double peak, indicating that it is associated with an $LP_{11}$ mode-locking state. To further verify the $LP_{11}$ mode-locking, the upper and lower lobes shown in Fig. 3(d) are measured separately, and the corresponding pulse trains and spectra are compared in Figs. 3(f-g). The pulse trains and spectra recorded from the two lobes are almost the same. The spectrum of the entire beam is slightly different from those of the two lobes, likely due to the existence of a few high-order modes. We note that the occurrence of beam auto-selection mode-locking depends on the coupling condition from the free space to the step-index MMF. Only in a certain coupling state can the transition be achieved. Occasionally, we further observed a bistable state, i.e., the state frequently transitions between multimode and single-mode mode-locking in a certain fixed cavity setup.

In an MMF cavity, both the nonlinear gain of active fiber and the nonlinear loss of SA induce nonlinear energy exchange among transverse-modes. The former comes from spatial gain competition, and the latter is illustrated in Fig. S4 [27]. The SA originates from nonlinear polarization rotation in our cavity, and its parameters are adjusted by the rotation of waveplates. Therefore, the observed transition of mode-locking states can be directly attributed to the adjustment of the parameters of the SA. Subsequently, the effect of SA interacts with other effects in the cavity, especially with the spatial gain competition, determining the output mode components.

We demonstrate it by simulating the STML states in an MMF cavity with a different SA parameter, saturation power $P_{sat}$. Under $P_{sat}$=12 kW (case 1), the stable STML outputs are presented in Figs. 4(a, b). In this state, mode 1 dominates, and there are mainly two modes, modes 1 and 6, with large walk-off that propagate in the MMFs. In this case, the intracavity evolution of the walk-off between these two modes is similar to that in Fig. 1(c). For case 2 with a $P_{sat}$ value of 36 kW, the corresponding intracavity evolution and stable output are given in Fig. 4(c-e) (See Movie S2 [38] for a comparison of the intercavity evolutions of these two cases with the same initial input). With increasing $P_{sat}$, mode 2 ($LP_{11}$) emerges, although other conditions are fixed. In this case, there is only one transverse-mode ($LP_{11}$) that propagates in both the step-index and graded-index MMFs, as shown in Fig. 4(e), which is significantly different from case 1. Note that after propagation through the SA, mode 2 splits into mode 2 plus a small part of mode 10, as shown in Figs. 4(d-e), likely because the profile of mode 10 resembles that of mode 2. The corresponding spatiotemporal intensity in Fig. 4(c) is dominated by the shape of the $LP_{11}$ mode. We note that, a similar STML state transition has been predicted in a paper published after the submission of this work [40].

In conclusion, the spatiotemporal self-organization of the multimode solitons is demonstrated in MMF lasers with large modal dispersion, the nonlinear dynamical process of which is quite different from that inside the graded-index MMF cavity. Moreover, a new spatiotemporal nonlinear dynamic, passive auto-selection of single-mode mode-locking, is reported and analyzed. SA plays critical roles in achieving STML, as well as in these spatiotemporal dynamics. From another perspective, for a multimode cavity with large modal dispersion, two mode-locking states are expected. One is multimode, assisted by the SA balancing the walk-off among different transverse-modes, and the other is single-mode (without suffering from the modal walk-off), which is achieved through the interactions between the SA and mode competition.

The introduction of active step-index MMFs into STML cavities provides new opportunities for research at both the scientific and technological levels. From a fundamental perspective, although the nonlinear dynamic process has been extensively studied in mode-locked single-mode lasers, it is still poorly understood in multimode cavities with immense spatiotemporal complexity. Our results expand the understanding of complex nonlinear spatiotemporal dynamics. In addition, the passive nonlinear auto-selection of mode appears to allow for a compromise between the benefits of multimode and single-mode systems. Furthermore, the STML lasers with large modal dispersion extend the explorable parameter space of the new spatiotemporal dynamics that can be achieved within them. In terms of technology, our work greatly broadens the design possibilities and provides convenient architectures for STML cavities because the active step-index MMFs are easy to fabricate and are commercially available. STML lasers with step-index multimode gain fibers are promising to be exploited as convenient platforms for the investigation of high-dimensional nonlinear science and as high-power, ultrashort optical pulse sources.

**Acknowledgements**

The authors would like to thank Dr. Logan G. Wright for helpful discussion and suggestion, and the help of numerical simulations. This work was partially supported by the National Key Scientific Instrument and Equipment Development Project of China under Grant 2014YQ510403, National Natural Science Foundation of China (51527901, 61975090), Fund of State Key Laboratory of IPOC (BUPT), P. R. China (No. IPOC2019ZZ02 and IPOC2020ZT02).

**Author contributions**

X.X. and Y.D. conceived the ideas. Y.D. conducted the experiments. X.X. performed the simulations. K.L., S.F. and X.X. helped performing the experiments. X.X., X.Z. and C.Y. supervised the project. Y.D. and X.X. co-wrote the manuscript. All authors commented on the manuscript.

†These authors contributed equally to this work.

*Corresponding authors. Email: xsxiao@bupt.edu.cn (X.X.); cxyang@mail.tsinghua.edu.cn (C.Y.)

**Figures**

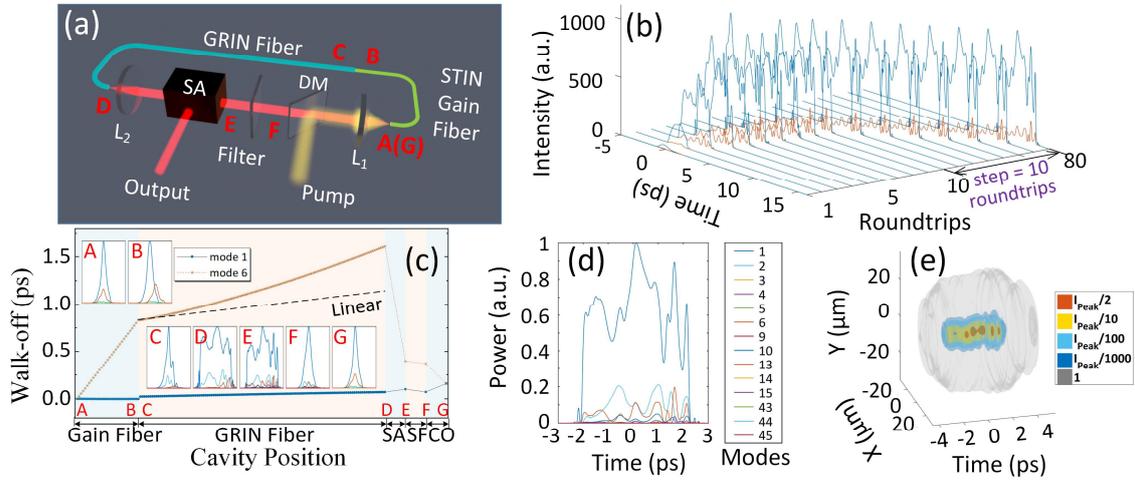

FIG. 1. Mode-resolved simulation of STML in an MMF laser with an active step-index MMF. (a) Cavity used in the simulation ($L_{1,2}$: lenses, DM: dichroic mirror). (b) Intercavity evolution of the output pulse. For clarity, only modes 1 and 6 are shown. (c) Intracavity evolution of the walk-off between modes 1 and 6 when stable STML is achieved. The data are calculated by the center of gravity for both modes relative to a reference frame moving with the linear group velocity of mode 1. Black dashed line: linear walk-off of mode 6 related to mode 1. Insets A-G: Temporal pulse shapes at different cavity positions labelled in (a) [A/B: input/output of the step-index (STIN) gain fiber, C/D: Input/output of the graded-index (GRIN) fiber, E: after SA, F: after spectral filtering (SF), G: input of the gain fiber via coupling (CO)]; The lengths of step-index and graded-index MMFs are 0.6 and 2.4 m, respectively. (d) Mode-resolved temporal output; (e) The corresponding spatiotemporal intensity of (d). $I_{peak}$: peak intensity of the multimode pulse.

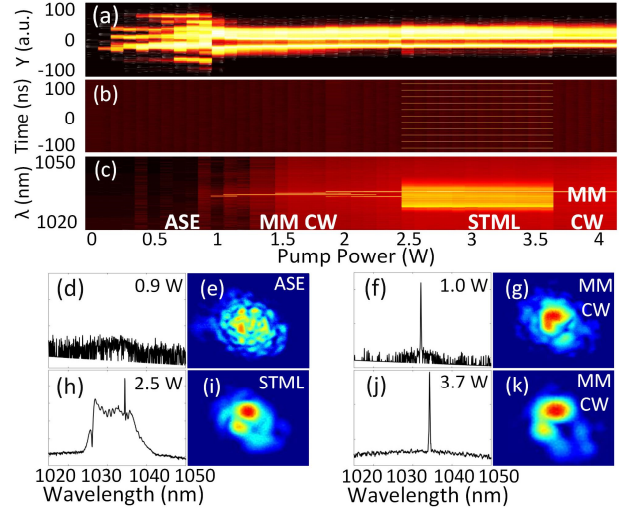

FIG. 2. Evolution of output with increasing pump power in the cavity. The evolutions of (a) the beam profile integrated over one dimension (the intensity is normalized for each step), (b) temporal output over a 200-ns span, and (c) spectrum. As the pump power increases, the field transitions from amplified spontaneous emission (ASE) [(d-e)] to multimode continuous-wave lasing (MM CW) [(f-g)] and then to STML [(h-i)] and eventually to MM CW [(j-k)]. The values of corresponding pump powers are labeled.

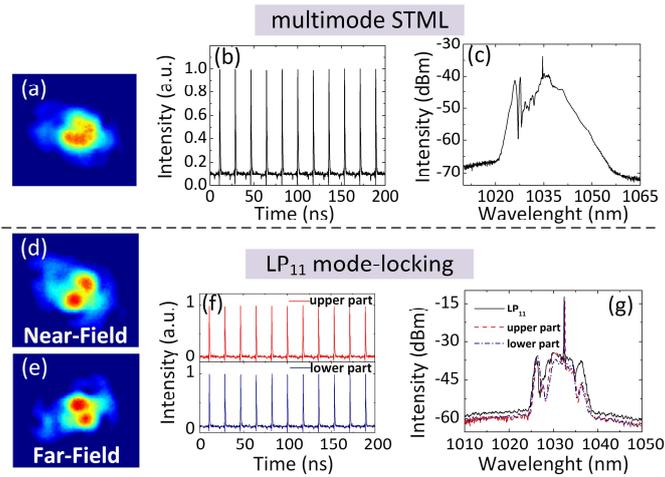

FIG. 3. Transition from multimode STML to single transverse-mode ($LP_{11}$) mode-locking in the same MMF cavity by adjusting the intracavity waveplate only. (a) Beam profile, (b) pulse train, and (c) spectrum of the output pulse train of the multimode STML state. (d-g) $LP_{11}$ mode-locking. Output beam profiles in the near field (d) and far field (e). (f) Pulse trains of the upper and lower parts of the beam profile in (d). (g) Spectra of the entire beam profile and the upper and lower parts of the beam profile in (d).

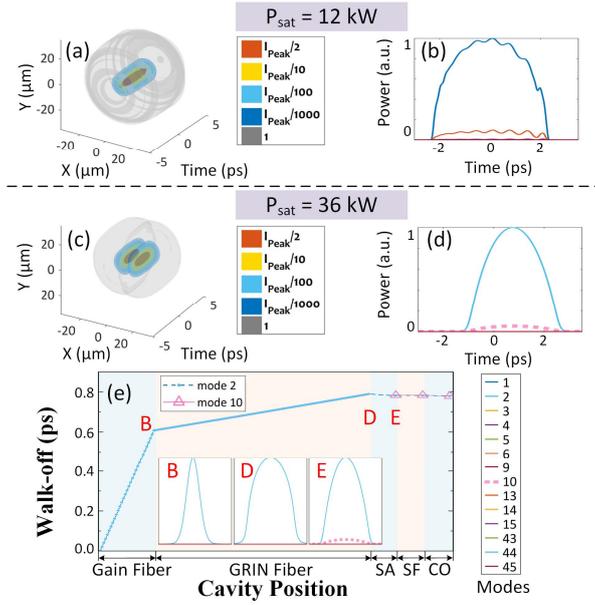

FIG. 4. Simulation of STML in an MMF laser with different SA saturation powers ($P_{sat}$), demonstrating the dependence of STML state on the SA parameter. (a), (c) Spatiotemporal intensity of and (b), (d) mode-resolved stable output pulses for (a-b) $P_{sat}$ = 12 kW (case 1) and (c-d) $P_{sat}$ = 36 kW (case2). (e) Intracavity evolution of the walk-off of mode 2 (and mode 10 after the SA) relative to a reference frame moving with the linear group velocity of fundamental mode when stable STML is achieved in case 2. Insets B, D-E: mode-resolved temporal pulse shape at the different cavity positions labelled in Fig. 1(a) [B/D: output of the step-index gain fiber/graded-index (GRIN) fiber, E: after the SA].

# Supplemental Material

The supplemental material including:

# Section S1. Overview of the methods for simulations and experiments

**Simulations.**

The cavity model is given in Fig. 1(a). The generalized multimode nonlinear Schrödinger equation [GMMNLSE, Eq. (S1)] is used for the simulation of nonlinear pulse propagation in passive MMFs, which is solved by a massively parallel algorithm [1,2]. The simulation parameters of MMFs are based on the fibers used in the experiments. The core diameter and numerical aperture (NA) of step-index (STIN) MMF are 20 μm and 0.08, respectively, and those of the graded-index (GRIN) MMF are 50 μm and 0.2, respectively, with a refractive index power of 2.08. In Figs. 1 and 4, the lengths of the STIN and GRIN MMFs are 0.6 and 2.4 m, respectively. The spatial modes, dispersions, and nonlinear coupling coefficients are calculated from the fiber specifications (some of these calculated parameters of the STIN MMF are shown in Fig. S1). All 6 modes of the STIN MMF are considered, and for the GRIN MMF, different mode combinations are considered.

To model the gain competition among the transverse modes, spatially dependent gain saturation is considered in the active MMFs. The gain spectrum has a Gaussian profile with a full-width-at-half-maximum (FWHM) of 40 nm. An instantaneous spatiotemporal SA model is used, the transmission function of which is applied to the full spatiotemporal field composed of all modes. A portion of the SA output is taken as the laser output. Then, a spectral filter with a Gaussian profile and FWHM of 10 nm is applied. We started the simulations at the input of the STIN gain fiber with small pulses in all 6 modes. Simulations with other cavity parameters and simulation setup (e.g., different combinations of modes and initial inputs) were tested, and we found that stable STML can be achieved for a wide variety of parameters. More details on the simulations can be found in section S2 and Refs. [1,2].

**Experiments.**

The cavity configuration is given in Fig. S5, where the STIN gain fiber is a Yb-doped MMF (Nufern LMA-YDF-20/125-9M) with a 20-μm core diameter and NA=0.08. The passive GRIN fiber is an OM4 fiber (YOFC Corp.) with a 50-μm core and NA=0.2. The lengths of the STIN and GRIN fibers are 0.4 m and 2.6 m, respectively. Other fiber lengths are also tested (see section S5.3). The free-space section includes waveplates, an isolator, a polarization beam splitter and a spectral filter (bandpass filter with a center wavelength of 1030 nm and FWHM of 10 nm). The fundamental repetition frequency of the cavity is 56.1 MHz, corresponding to a 17.8 ns roundtrip. The gain fiber is pumped by a 976-nm laser diode. Self-starting STML occurs with appropriate cavity setup. More information about the cavity configuration and field measurements can be found in the following (see Fig. S5 and section S4) and Refs. [1,3,4].

## Section S2. Additional simulation details

The equation applied in the simulation of pulse propagation in the passive MMFs is the generalized multimode nonlinear Schrödinger equation (GMMNLSE) [1,2], which is widely used in numerical research related to MMFs. The GMMNLSE can be written as follows:

$$\partial_z A_p(z,t) = i(\beta_0^{(p)} - \Re[\beta_0^{(0)}])A_p - (\beta_1^{(p)} - \Re[\beta_1^{(0)}])\frac{\partial A_p}{\partial t} + i\sum_{n\geq 2}\frac{\beta_n^{(p)}}{n!}(i\frac{\partial}{\partial t})^n A_p + \\ i\frac{n_2\omega_0}{c}(1+\frac{i}{\omega_0}\partial t)\sum_{l,m,n}\left\{(1-f_R)S^K_{plmn}A_l A_m A_n^* + f_R A_l S^R_{plmn}\int_{-\infty}^{t}d\tau A_m(z,t-\tau)A_n^*(z,t-\tau)h_R(\tau)\right\}$$ (S1)

where $A_p(z,t)$ is the electric field temporal envelope for spatial mode $p$, $\beta_n^{(p)}$ is the $n$th order dispersion for mode $p$, and $S^K_{plmn}$ and $S^R_{plmn}$ are the nonlinear coupling coefficients for the Kerr and Raman effects, respectively, with $p$, $l$, $m$, and $n$ representing the numbers of spatial modes. $f_R$ is the Raman contribution to the Kerr effect, $h_R$ is the Raman response of the fiber, $\Re$ denotes the real part only, and $n_2$ is the nonlinear index of refraction. In our simulations, the center wavelength is set to 1030 nm, and $n_2 = 2.3 \times 10^{-20}$ m$^2$W$^{-1}$; additionally, the high-order (3$^{rd}$, 4$^{th}$,...) dispersion, Raman ($f_R = 0$), and shock terms are neglected. The dispersion and nonlinear coupling coefficients are calculated from the fiber specifications. We started the simulations at the input of the STIN gain fiber with evenly distributed Gaussian pulses in all 6 modes, for which the total energy is 0.01 nJ and the FWHM duration is 1 ps. For the STIN MMF, $\Delta\beta_1^{(p)} = \beta_1^{(p)} - \Re[\beta_1^{(0)}]$ of the 6 modes are shown in Fig. S1(b), and the values of the group-velocity dispersion ($\beta_2^{(p)}$) of the 6 modes are in the range of 18-23 fs$^2$/mm; for the GRIN MMF, $\Delta\beta_1^{(p)}$ of modes 2-6 are 0.07, 0.07, 0.15, 0.15, and 0.15 fs/mm, respectively, and $\beta_2^{(p)}$ are approximately 18 fs$^2$/mm. All 6 modes of the STIN MMF are considered, and for the GRIN MMF, approximately 120 modes can be guided at a wavelength of 1030 nm. Although a massively parallel algorithm is used to accelerate the calculation, the simulation will become very slow for more than 20 modes. So we have to choose part of the modes in GRIN MMF. Many simulations with different mode combinations were tested. Based on the simulation results, the modes important for the simulations were selected. As shown in Figs. 1 and 4, 14 modes with numbers of (1-6, 9, 10, 13-15, 43-45) were considered. Simulations with more modes were tested, with similar numerical results.

For the active STIN MMF, optical gain with a finite bandwidth and gain saturation is considered, in addition to the GMMNLSE. To model the gain competition among modes, the spatial dependence of gain saturation is considered. The model of the gain considering the spatial dependence of gain saturation is as follows [5]:

$$\frac{\partial A(x,y,z,\omega)}{\partial z} = \frac{1}{1+\int dt |A(x,y,z,\omega)|^2 / I_{sat}} \frac{g}{2} f(\omega) A(x,y,z,\omega)$$

where $A(x,y,z,\omega) = \sum_n F_n(x,y) A_n(\omega,z)$ with $F_n(x,y)$ be the transverse mode profile of the *n*th mode, $I_{sat}$ is the saturation time-integrated intensity, $g$ is the small signal gain, and $f(\omega)$ reflects the gain bandwidth. The small signal gain $g$ is 33 (30) dB in Fig. 1 (4), and the gain spectrum with a Gaussian profile and an FWHM of 40 nm is considered. The saturation energy of the gain fiber is 8 nJ, i.e., $I_{sat}$ =8 nJ/$A_{eff}$, where $A_{eff}$ is the effective mode area of the fundamental mode.

The transmission function of the SA with a spatiotemporal effect, as applied to the full spatiotemporal field *A(x,y,z)*, is as follows [1]:

$$A_{out}(x,y,z) = A_{in}(x,y,z) \sqrt{1 - \frac{q}{1+\frac{|A(x,y,z)|^2}{I_{sat}}}}$$

where *A(x,y,z)* is the summation of all modes, q is modulation depth and $I_{sat}$ is the saturation intensity of the SA. The *q* value of the SA is set to 1, and $I_{sat}$ of the SA is calculated by $P_{sat}$ /$A_{eff}$, where $A_{eff}$ is the effective mode area of the fundamental mode. The saturation power $P_{sat}$ is 16 kW in Fig. 1 and 12 or 36 kW in Fig. 4. After SA transmission, a lumped loss of 80% is considered, including the loss due to the laser output. Then, a spectral filter centerd at 1030 nm with a Gaussian profile and a FWHM of 10 nm is applied to the field.

There are two coupling relations in the cavity model: the coupling from the STIN to the GRIN MMFs, and that from the GRIN to the STIN MMFs. We model the process of mode-resolved coupling by $A_{out}$=M·$A_{in}$, where $A_{in}$ ($A_{out}$) is an *n*x1 (*m*x1) vector with each element being the field of a mode before (after) coupling and *n* (*m*) is the number of modes. M is an *m*x*n* matrix of coupling coefficients that reflects the excitation of modes in the MMFs due to optical coupling. For Fig. 1, the coupling matrix M from the STIN gain fiber to passive GRIN MMF is set with M(1,1)=$\sqrt{3}$/2, M(1,6)=M(2,1)=$\sqrt{2}$/4, M(2,2)=1/2, and M(3,3)= M(4,4)= M(5,5)= M(6,6)=1/3. The other components of M are zero. For Fig. 4, M(1,1)= 0.95, M(2,2)=1, and M(3,3)= M(4,4)= M(5,5)= M(6,6)=1/3, and the other components of M are zero. For simplicity, the coupling matrix from the GRIN MMF output to the STIN gain fiber is set as the transposition of M.

## Section S3. Additional numerical results

**S3.1 Comparison between graded- and step-index MMFs**

GRIN fibers are usually designed to exhibit small modal dispersion for using in telecommunications with small signal degradation induced by dispersion. While STIN fibers usually exhibit much larger modal dispersion. For these two types of fibers with similar fiber specifications, the modal dispersion (walk-off among different modes) of STIN MMFs is approximately one order of magnitude larger than that of GRIN MMFs. Herein the GRIN and STIN MMFs are compared briefly. There is a large difference in the dispersion, as well as nonlinearity, characteristics between STIN and GRIN MMFs [2,6]. We illustrate this difference in Fig. S1 by comparing the coefficients of these two fibers with similar specifications. The parameters of STIN MMFs are set in accordance with the commercial fibers used in the experiments (Nufern LMA-YDF-20/125-9M with a 20-μm core diameter, NA=0.08), supporting up to 6 modes, as illustrated in Fig. S1(a). As seen from the top portion of Fig. S1(b), modes 4-6 of the GRIN fibers form a mode group with almost the same propagation constant ($\beta_0$), while mode 6 of the STIN fibers has different $\beta_0$ values than other modes. The characteristics of $\beta_0$ has an influence on the nonlinear multimode interactions [2]. More importantly, the modal dispersion shows quite different behaviors between those two fiber types, as shown in the bottom part of Fig. S1(b). The group velocities of the high-order modes (HOMs) related to the fundamental mode ($\Delta\beta_1$) in STIN fibers are much larger than those in GRIN fibers, for which the group velocities of all 6 modes are almost the same. The large walk-off among the spatial modes in STIN fibers requires more effort to achieve compensation. Thus it is difficult to achieve the self-organization of localized multimode pulses in cavity with STIN fibers compared to that using GRIN fibers. Besides the dispersion, as shown in Fig. S1(c), the nonlinear coefficients ($S^K_{plmn}$) of HOMs in STIN fibers are larger than those in GRIN fibers, in terms of both self-phase modulation and cross-phase modulation among modes. The variations in the nonlinear coefficients of the spatial modes in GRIN fibers are wider than those in STIN fibers.

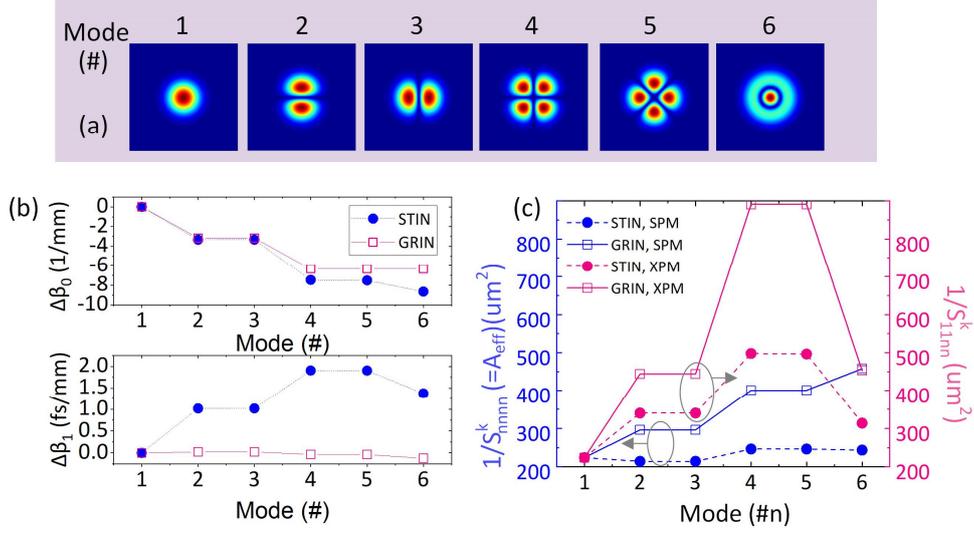

Fig. S1. Comparison of the dispersion and nonlinearity between STIN and GRIN MMFs, with identical index contrast and effective mode areas ($A_{\text{eff}}$) for the fundamental mode (mode 1). (a) Illustration of the spatial modes 1-6 in MMFs. (b) Propagation constant ($\Delta\beta_0$) and group velocity ($\Delta\beta_1$) of modes related to mode 1. (c) Coefficients reflecting the self-phase modulation (SPM, effective mode area $A_{\text{eff}}=1/S^{K}_{nnnn}$) of each mode and the cross-phase modulation (XPM, $1/S^{K}_{11nn}$) between modes 1 and n. $S^{K}_{p,l,m,n}$ is the nonlinear coupling coefficient for the Kerr effect, with $p$, $l$, $m$, and $n$ representing the numbers of spatial modes [2].

### S3.2 The effect of saturable absorber

For the simulation case of Fig. 1, Fig. S2 below shows the input and output of SA, which is extracted from the Movie S1. In order to show the effect of SA on the pulse, the mode-resolved temporal pulse and spectra, as well as the corresponding spatiotemporal intensity are presented.

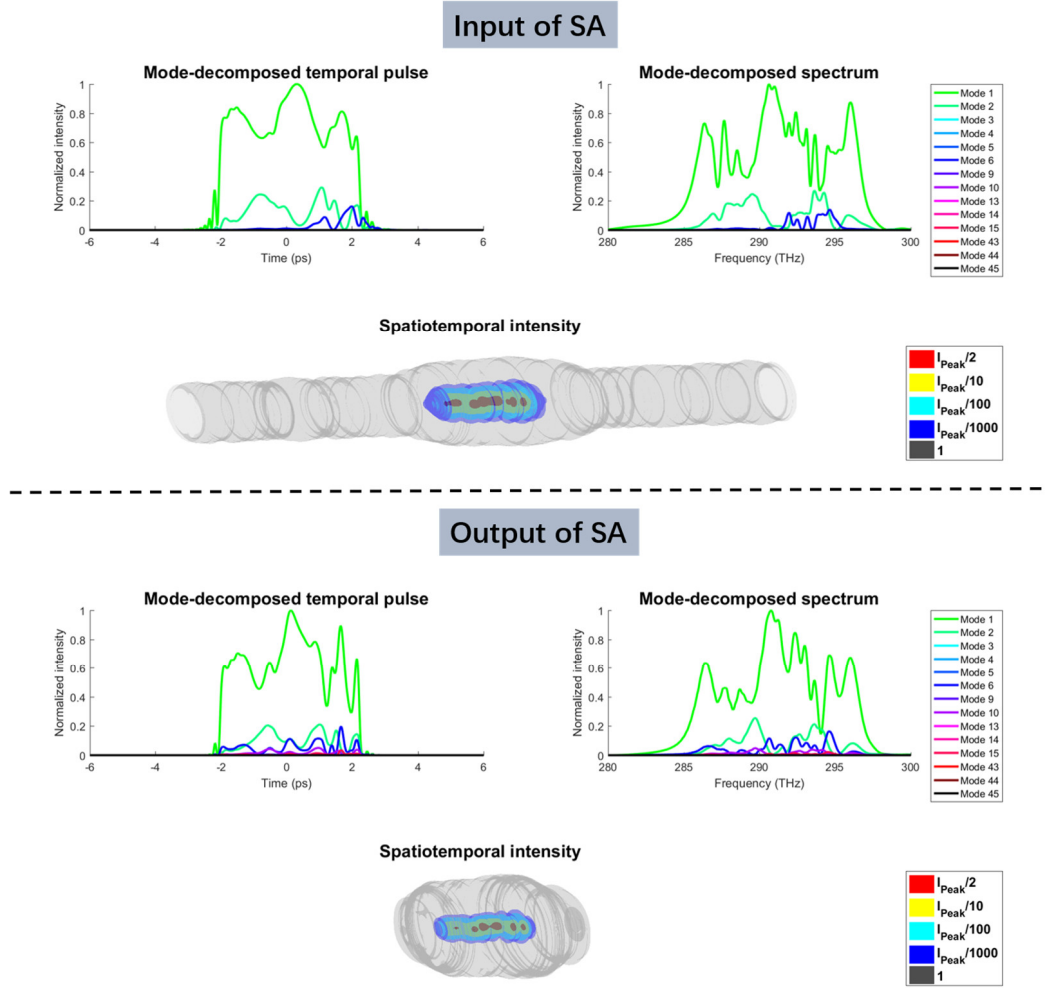

**Fig. S2. Input and output of SA.** It is the simulation results of the case in Fig. 1(c), and is extracted from Movie S1.

## S3.3 Comparison of the intracavity evolutions of the modal walk-off in the MMF cavities with large and small modal dispersion

Firstly, we explain the unusual walk-off in the GRIN fiber shown in Fig. 1(c), which is different from that shown in the following Fig. S3 with small modal dispersion. We infer that the larger walk-off in GRIN fiber is caused by fiber nonlinearity, which will be greatly affected by linear effects such as dispersion. In the case of Fig. 1(c), there is a large walk-off between modes 1 and 6 due to the large modal dispersion of STIN fiber, thus mode 6 overlaps with the tail of mode 1 for a long time (almost throughout the two segments of MMFs), as shown in the Movie S1 and insets of Fig. 1(c). The pulse tail of mode 1 is steep and the slope of the tail is large. This results in a large frequency shift to mode 6 through the nonlinear cross-phase modulation (XPM) from mode 1. Subsequently, the intermodal XPM induced frequency shift causes additional walk-off through chromatic dispersion, besides the linear walk-off caused by modal dispersion. In short, the unusual modal walk-off in GRIN fiber is originated from the large modal dispersion of STIN fiber. This conclusion is validated by the comparison between Fig. 1(c) and Fig. S3.

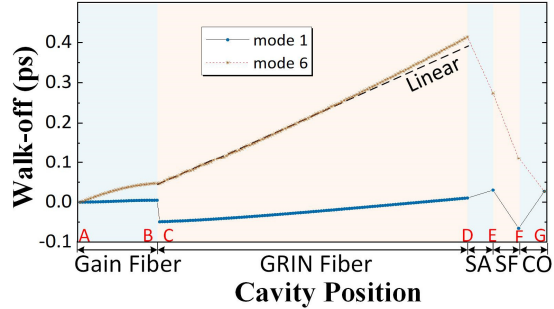

**Fig. S3. Intracavity evolution of the walk-off between modes 1 and 6 in the cavity with small modal dispersion.** The description of this figure refers to Fig. 1(c). Black dashed line: linear walk-off of mode 6 related to mode 1.

One can find the difference of intracavity evolutions between our cavity and the GRIN MMF cavity by comparing our results to these of Ref. [1] (e.g., comparing Fig. 1(c) in this article to Figs. S4D, S8D, S12D in the Supplementary Materials of Ref. [1]). In order to further show the different evolutions inside MMF cavities with large and small modal dispersion, simulations for an MMF cavity with small modal dispersion are conducted. In this cavity, the dispersion parameters of all the 6 modes in the active STIN fiber are set to the same as these of the 6 lowest-order modes in the passive OM4 fiber. Other parameters are the same as the case in Fig. 1. The intracavity evolution of the walk-off between modes 1 and 6 for this MMF cavity is given in Fig. S3, which can be compared to Fig. 1(c) with large modal dispersion. As seen from Fig. S3, due to the small modal dispersion in gain fiber, the walk-off between modes 1 and 6 is small. Thus the frequency shift of mode 6 resulting from the cross-phase modulation of mode 1 is much smaller than the case of Fig. 1. Subsequently, in the following GRIN fiber, the walk-off is almost the same as the linear walk-off, which is calculated by $(\beta_1^{(6)} - \beta_1^{(1)})*z$. Here $\beta_1^{(1)}$ and $\beta_1^{(6)}$ are the group velocities of modes 1 and 6 calculated based on the fiber specifications, and $z$ is the propagation distance. At the output of GRIN fiber, the walk-off is much smaller than that of Fig. 1(c). Eventually, the small walk-off is compensated for by the combination of SA and filtering.

**S3.4 Energy exchange among transverse modes caused by saturable absorber**

In the MMF cavity investigated, nonlinear gain/loss mainly results from SA and gain fiber, which will cause energy exchange among transverse modes (energy exchange caused by fiber nonlinearity is found very small in the cavity). In order to show the effect of SA, Fig. S4 illustrates the influence of SA on the relative energy of modes 1 and 2. The simulation cases are these of Fig. 4. As seen from Fig. S4, SA will change the relative energy of modes 1 and 2, and for different $P_{sat}$ the change is different. For roundtrip #1, the inputs of SA are the same for the two cases. However, the outputs, including the energy ratio of modes 1 to 2, are different due to different SA's parameter ($P_{sat}$). Enhanced by the spatial gain competition in active fiber, the difference between the two cases becomes more visible.

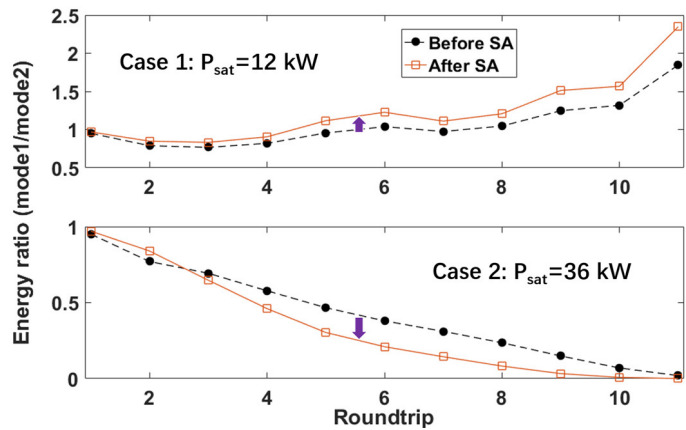

**Fig. S4. Influence of SA on the energy ratio of modes 1 to 2.** The corresponding intercavity evolutions can refer to Movie S2.

## Section S4. Additional details of the experimental setup

### S4.1 Experimental setup

The experimental scheme is shown in Fig. S5. The scheme is similar to that in our previous work [3,4], except that the gain fiber we adopted here is a STIN MMF (Nufern LMA-YDF-20/125-9M with a 20-μm core diameter) to induce more HOMs. The STIN gain fiber is pumped using a 976-nm laser diode with an 8-W maximum pump power through a dichroic mirror. To minimize the insertion loss, the STIN fiber is spliced to the GRIN fiber without offset. The lengths of the gain fiber and OM4 are 0.6 m and 2.4 m, respectively. Both fibers are placed loosely to prevent HOM leakage. The laser is passively mode-locked by nonlinear polarization rotation (NPR) achieved using two quarter-wave plates (QWP), one half-wave plate (HWP) and one polarization beam splitter (PBS). Then, the light is ejected by the PBS as output. An isolator (ISO) is added after the PBS to ensure a unidirectional propagation. A 10-nm bandpass spectral filter (SF) at a 1030-nm center wavelength is used for the self-consistency of the intracavity pulses and to compensate for the normal group-velocity dispersion of both MMFs. The light is reflected back to the STIN gain fiber by a mirror. With appropriate cavity conditions (sets of waveplates, mirror alignments, and pump power), STML is readily obtained. By tuning the mirrors, different coupling conditions result in many different spatiotemporal outputs.

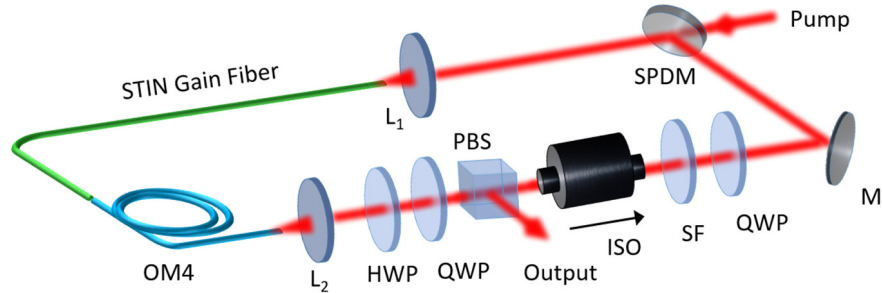

**Fig. S5. Experimental setup of the multimode cavity with a STIN fiber.** SPDM, short pass dichroic mirror; $L_1$ and $L_2$, lenses; STIN gain fiber, step-index gain fiber; OM4, passive GRIN multimode fiber; HWP, half-wave plate; QWP, quarter-wave plate; PBS, polarization beam splitter; ISO, isolator; SF, spectral filter; M, mirrors.

### S4.2 Field measurements

The output is directly measured by an optical spectrum analyzer (OSA) with a 0.06-nm resolution, a 5-GHz photodetector with an 8-GHz real-time oscilloscope and a radio frequency (RF) signal analyzer, and a beam profiler. A second-harmonic autocorrelator with a 150-ps measurement range is used to measure the autocorrelation (AC) signals.

Spectral-filtering and spatial sampling measurements are adopted here to further verify STML [1,3,4]. For spectral-filtering, a tunable bandpass spectral filter is employed after the output. For spatial sampling, part of the beam is collected by an MMF with a 50-μm core diameter without a focusing lens. By adjusting the relative position of the sampler, we can obtain various spatial parts of the beam. Details can be found in Refs. [4] and [1].

## Section S5. More proofs of STML

### S5.1 Typical output of STML states

Stable STML with uniform output pulses can be readily achieved and self-started. With the measurements described in Section S4.2, a typical result with a 4-W pump power are illustrated in Fig. S6. Fig. S6(a) presents a pulse train with a uniform intensity and an interval of 17.8 ns, corresponding to a 56.1-MHz fundamental repetition rate. Fig. S6(b) shows the RF spectrum with a resolution bandwidth of 1 Hz over a 2.5-kHz span. The inset of Fig. S6(b) is the RF spectrum over a 1-GHz span, proving a uniform intensity pattern. The AC trace in Fig. S6(c) indicates a pulse duration of 4.1 ps, assuming a profile of Gaussian shape. Spectral-filtering and spatial sampling measurements are applied to further confirm STML (see section S4.2 above and Refs. [1,3,4], especially Fig. 1 and Section 3.1 of Ref. [4], for detailed information).

Using a tunable spectral filter, various combinations of transverse modes can be selected and measured. As shown in Fig. S6(d), three different spectrum regions (i.e., different transverse mode components) are selected (dashed curves labelled by f, g, and h) from the whole output spectra (solid curve). The corresponding beam profiles for these sliced spectra are different, as shown in Figs. S6(f)-(h), while the corresponding RF spectra overlap, as shown in Fig. S6(i). This indicates that the transverse modes are synchronously locked.

The output is also spatially-sampled at two different positions, S1 and S2. The optical spectra and the corresponding pulse trains and RF spectra of the two spatially-sampled outputs are shown in Figs. S6(j)-(l), respectively. Similar to spectral-filtering described above, the optical spectra are distinct to each other, indicating that the sampled transverse mode components are different and confirming the multimode spatiotemporal structure of the output. Though the transverse mode components sampled at S1 and S2 are different, the RF spectra are almost the same, confirming the STML.

These results further suggest that STML indeed occurs in the MMF laser cavity. In the absence of a temperature-controlled laboratory environment, the STML operation could last for several hours.

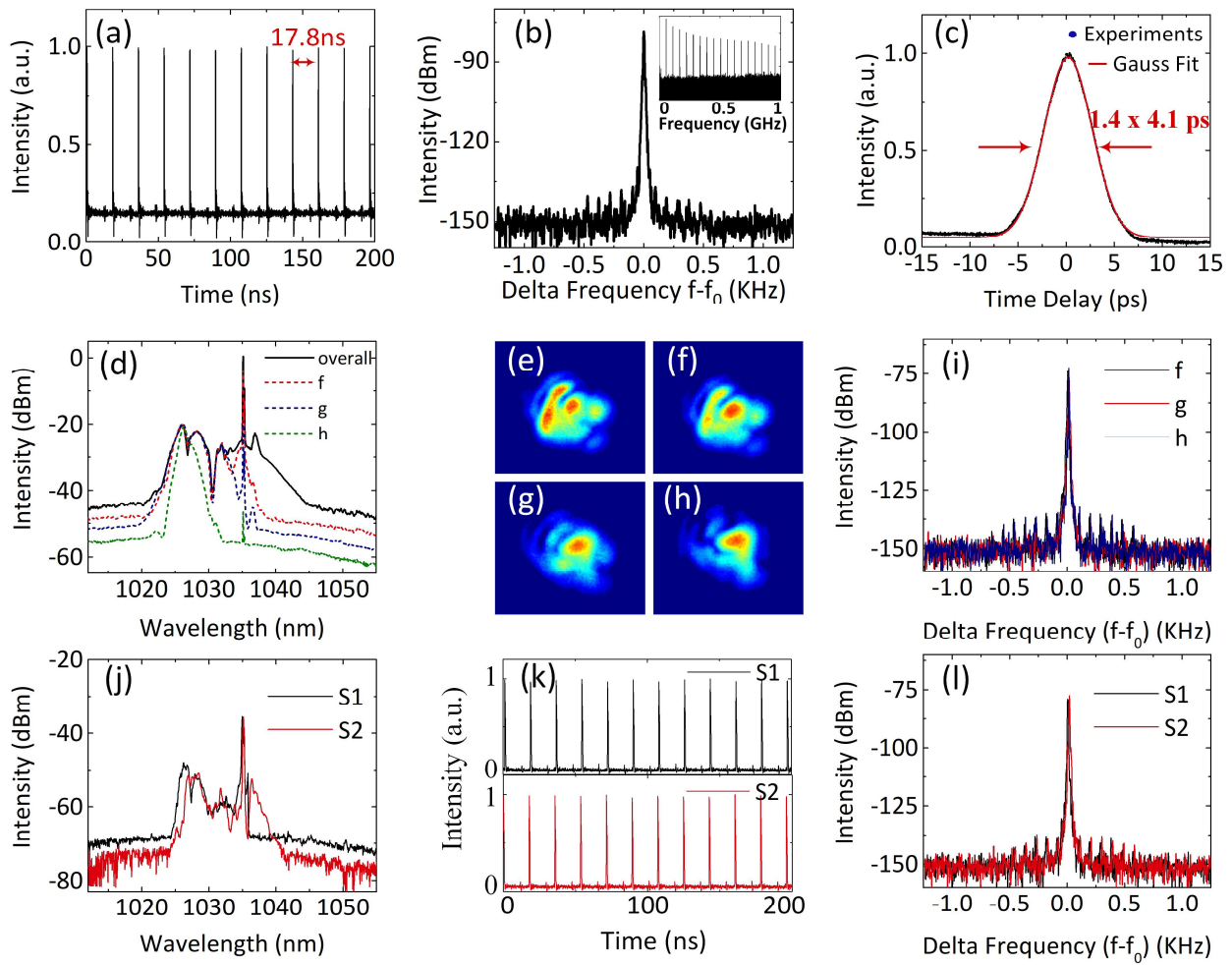

**Fig. S6. Typical STML state in the MMF laser cavity.** (a) Pulse train; (b) RF spectrum; (c) AC trace; (d) Spectra of the entire output (solid curve, corresponding beam profile shown in (e)) and the spectral-filtered outputs (dashed curves f, g, h) with the corresponding beam profiles and RF spectra shown in (f-h) and (i); (j) Spectra sampled at two different positions (denoted as S1 and S2); (k) and (l) are the corresponding pulse trains and RF spectra of S1 and S2, respectively.

Usually a continuous-wave component exists in the output. However, by carefully tuning the waveplates, the continuous-wave component can be suppressed, e.g., as shown in Fig. S7.

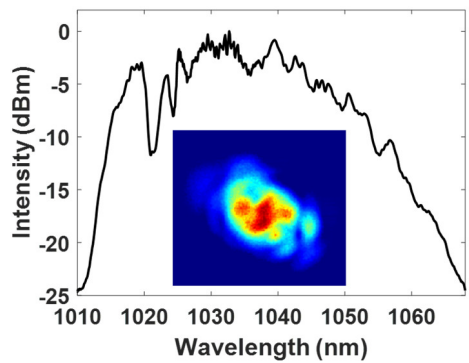

**Fig. S7. Another typical STML state.** Optical spectrum with beam profile inserted.

## S5.2 Single-shot spectra by dispersive Fourier transformation technique

The dispersive Fourier transformation (DFT) technique [7] is utilized here to further discuss the STML state. The stable mode-locking operation in different scenarios is obtained by different sets of waveplates at a 4-W pump power, and the beam profile of another STML state is shown in Fig. S8(a). With spatial sampling, we sample the beam at two different positions and then implement propagation through a 12.24-km single-mode fiber to stretch the pulse mapped to the temporal waveform. The real-time signals are detected by two high-speed photodetectors and an oscilloscope. For both sampled outputs, we record 100,000 roundtrip evolutions of the real-time spectra depicted in Figs. S8(b,d). Fig. S8(c,e) shows the corresponding exemplary single-shot spectrum. This result also confirms the stable STML state.

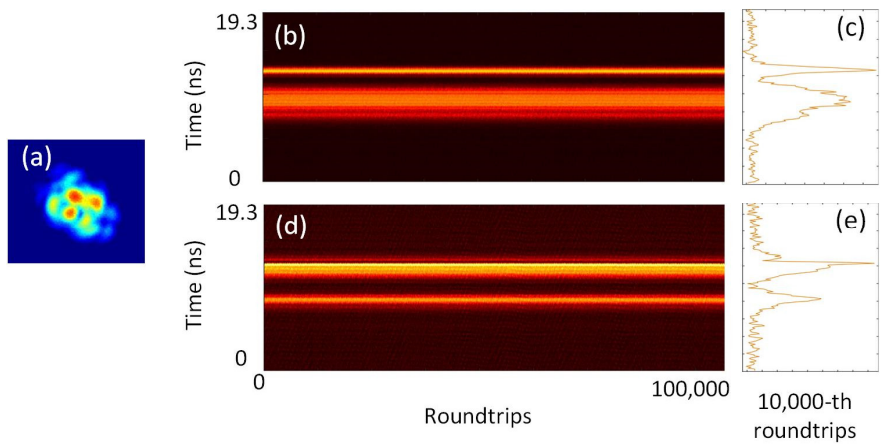

**Fig. S8. DFT results of a stable STML state in the MMF laser cavity.** (a) Output beam profile of the state. The beam is spatial sampled at two different positions, the DFT results of which are shown in (b) and (d), respectively. (b) and (d) are the real-time spectral evolution dynamics during the stable mode-locking state; (c) and (e) are the exemplary single-shot spectra at the 10,000th roundtrip, corresponding to (b) and (d), respectively.

In mode-locked single-mode fiber lasers, the DFT technique was usually used to investigate and visualize the buildup dynamics process [8,9]. In the further investigations of our MMF lasers, we measured the spatially-sampled buildup dynamics of the STML state by DFT technique. We found that, the initial oscillation processes of different spatial mode components are different. However, when the mode-locking is achieved, the single-shot spectra evolution of steady states are similar to Fig. S8. This preliminary investigation also indicates the STML of the MMF laser.

## S5.3 STML in cavity with other fiber lengths

We experimentally tested many cavities with various lengths of the active STIN MMF and passive GRIN

MMF used above. We achieved multimode STML in the cavities with short (~0.35 m) STIN gain fibers and found that the length of passive GRIN fibers should not be too short (usually > 1 m). For long (~1 m) STIN gain fibers, we cannot achieve STML, even for GRIN fibers of optimized length.

Here, we give measurement results of a stable STML state in a cavity with 0.6-m STIN gain fiber and 1.4-m GRIN fiber, as shown in Fig. S9. The pulse interval is 12.9 ns shown in Fig. S9(a), and corresponding optical and RF spectrum is presented in Figs. S9(b,c) with beam profile inserted.

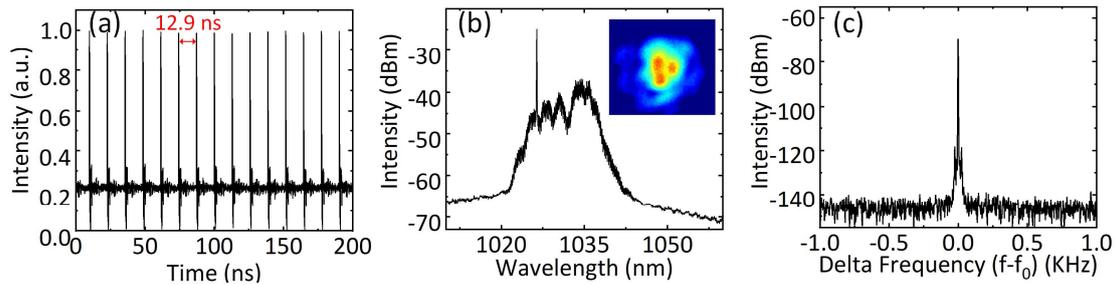

**Fig. S9. STML state in the cavity with 0.6-m active STIN fiber and 1.4-m GRIN fiber.** (a) pulse train; (b) Optical spectrum (inset: beam profile); (c) RF spectrum.

## S5.4 Coherence measurement

Go one step further to verity STML, we carried out interference experiments for the output from STML states. The interference setup is shown in Fig. S10. Due to the broadband spectrum of the multimode pulse, spectral interference [10] (which reflects the coherence) between different transverse components of a same multimode pulse was observed. For the STML state in the multimode laser, we extracted different time traces from beam profile with assistant of two samplers. Here the samplers are two multimode fibers with high numerical aperture, in order to collect enough power to observe the interference. For interference purpose, the two signals were emitted into a Mach-Zehnder-type interferometer with equal arm lengths. Half-wave plates placed in both arms, along with the polarizer, could adjust the light powers. The arm length was carefully adjusted by a tunable time delay device (i.e., pair of mirrors sitting on a moving stage). The two pulses from the two arms were combined by a beam splitter, then the interference signal was able to be detected by an optical spectrum analyser (Agilgent 86142B) with 0.06 nm resolution.

One of the sampling points in beam profile was fixed in the center. In the meantime, the other sampling point can be moved from edge to center through a 3-D stage. In this case, signal sampled from the center was dominated by fundamental mode, while signal sampled from the edge was dominated by higher order modes (HOMs). Spectra output from the interferometer in two different STML states were shown in Figs.

S11(a) and (b), respectively.

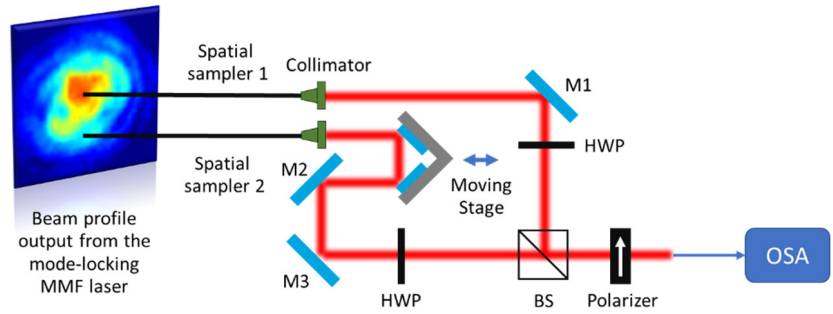

**Fig. S10. Setup for observing the coherence of transverse modes in the mode-locking state of the MMF laser.** M1-3, mirrors; HWP, half-wave plate; BS, beam splitter; OSA, optical spectrum analyzer.

At the output of the interferometer, when the lengths of two arms were carefully adjusted, spectral interferograms were observed (the spectral fringes in Fig. S11), confirming the coherence between different points of the spatial profile, thus validating the coherence between different transverse modes in our mode-locking states.

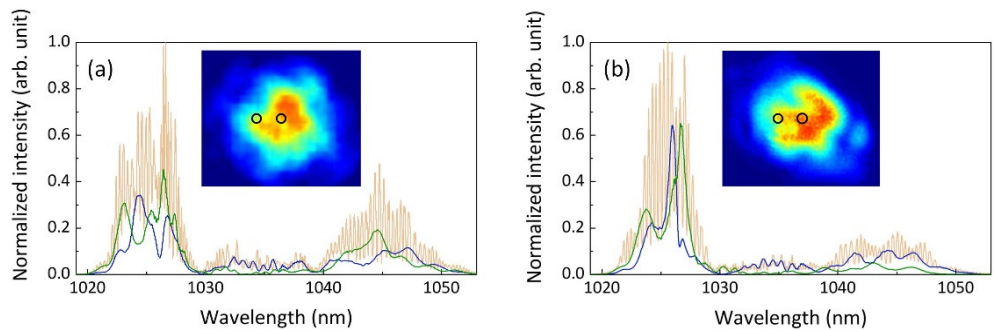

**Fig. S11. Spectra output from the interferometer for two typical STML states (a, b).** The beam profiles and the sampling positions (black circles) are given in the inset images. Spectral interferograms are depicted with orange line. Sampled spectra are also presented with green line (sampled from center) and blue line (sampled from edge), measured by blocking the other path of the interferometer.

What needs to be pointed out is that the modulation depth lower than 100% due to the following reasons: (1) The spectra of the two pulses are different. The spectra are broadband, and we can only adjust average intensity while the spectral intensities of the two pulses are not identical at most wavelength. This will weaken the modulation depth. (2) The non-ideal interferometer. The possible unideal factors include that (a) the spatial samplers, i.e. multimode fibers, degrade the coherence, (b) the bandwidth resolution limitation of the optical spectrum analyzer will also decrease the modulation depth [11], (c) the instability of Mach-Zehnder interferometer (e.g., the fluctuation of the two arms). In order to estimate the coherence, we conducted a comparison experiment. The spectral interferogram of a mode-locked single-mode fiber laser was measured by using the same setup above (Fig. S10 with equal powers of the two arms). In principle, the coherence of two points in the transverse profile of a single-transverse mode laser is perfect. However, due to the non-ideal interferometer, the modulation depth of the spectral fringe is also lower than 100%, as shown in Fig. S12(b).

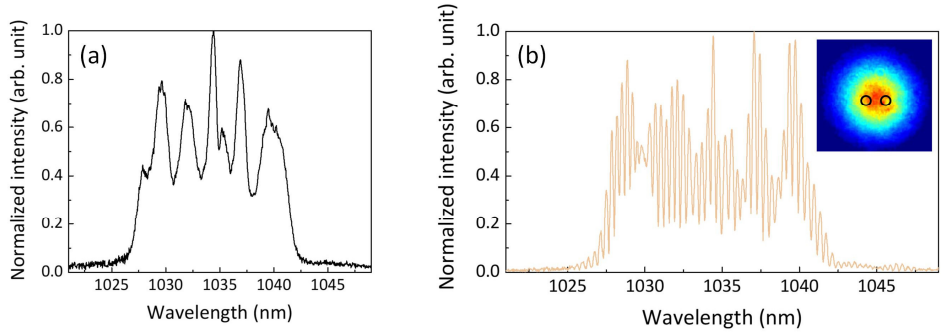

**Fig. S12. With the input from a mode-locked single-mode fiber laser (a), the measured spectrum output from the interferometer (b).** The beam profile and the sampling positions (black circles) are given in the inset image of (b).

We quantify the modulation depth of the interference spectra by calculating the average fringe visibility throughout the spectral bandwidth. The fringe visibility, $V$, is defined as

$$V = (I_{max} - I_{min})/(I_{max} + I_{min}),$$

where $I_{max}$ and $I_{min}$ are the maxima and adjacent minima of the spectral fringes, respectively [10]. The calculated average $V$ is 47.2% for the case of single-mode fiber laser (Fig. S12(b)); and for the STML states shown in Fig. S11(a) and (b), the average $V$ are 44.3% and 43.8%, respectively. The average fringe visibility of STML state is close to that of the mode-locked single-mode fiber laser, which indicates the good coherence in the STML state.

In all, by observing a modulated spectrum between pulses sampled from different points of the spatial profile (one at the center and the other at the edge), we prove the STML state in our multimode cavity.

## Description of Additional Supplementary Files (Movies S1 and S2)

**Movie S1. Intracavity evolution of the pulse in the case of Fig. 1.** This movie is the animated version of Fig. 1 with the intracavity evolution of the mode-decomposed temporal pulse, spectrum and the full spatiotemporal intensity of pulse. The intracavity step is 0.02 m for the STIN fiber and 0.08 m for the GRIN fiber.

**Movie S2. Intercavity evolutions of the pulse in the two cases of Fig. 4.** This movie is the animated version of intercavity evolutions in the two cases of Fig. 4 with the evolution of 30 roundtrips at $P_{sat}$=12 kW (case 1) and $P_{sat}$=36 kW (case 2). The movie includes a mode-decomposed temporal pulse, spectrum, the beam profile and the full spatiotemporal intensity of pulse. The initial inputs for the simulations of both cases are the same.